\documentclass[twocolumn]{revtex4}
\usepackage[usenames,dvipsnames]{color}
\usepackage{graphicx}
\usepackage{subfigure}
\usepackage{amssymb,amsmath,tikz}

\newcommand{\ket}[1]{\left|#1\right\rangle}
\newcommand{\bra}[1]{\left\langle#1\right|}
\newcommand{\mone}{M_{1\phi}}
\newcommand{\mtwo}{M_{2\phi}}

\begin{document}
\title{Information criteria for efficient quantum state estimation}

\author{J. O. S. Yin and S. J. van Enk}
\affiliation{Oregon Center for Optics, Department of Physics\\
University of Oregon\\
Eugene, OR 97403}
\date{\today}

\begin{abstract}
Recently several more efficient versions of quantum state tomography
have been proposed, with the purpose of making tomography feasible even for many-qubit states. The number of state parameters to be estimated is reduced by tentatively introducing certain simplifying assumptions on the form of the quantum state, and subsequently using the data to rigorously verify these assumptions. The simplifying assumptions considered so far were (i) the state can be well approximated to be of low rank, or (ii)
the state can be well approximated as a matrix product state.
We add one more method in that same spirit: we allow in principle any model for the state, using any (small) number of parameters (which can, e.g., be chosen to have a clear physical meaning), and the data are used to verify the model.
The proof that this method is valid
cannot be as strict as in above-mentioned cases, but is based on well-established statistical methods that go under the name of ``information criteria.'' We exploit here, in particular, the Akaike Information Criterion (AIC). We illustrate the method by simulating experiments on (noisy) Dicke states.
\end{abstract}
\maketitle

\section{Introduction}
Quantum state estimation \cite{PR2004b,H1997,B-K2010} remains one of the hot topics in the field of quantum information processing.  The  hope to recover each element in the density matrix, however, is impeded by the exponential growth of the number of matrix elements with the number of qubits, and the concomitant exponential growth in time and memory required to compute and store the density matrix. The task can become intimidating when 14 qubits are involved \cite{MSBCNCHHHB2010}, and so efforts have been made to simplify quantum state tomography.  One such effort focused on states that have high purity  \cite{GLFBE2010} so that the size of the state space shrinks significantly (from ${\cal O}(D^2)$ to ${\cal O}(D)$ for a system described by a $D$ dimensional Hilbert space). Given that the measurement record is used to verify the assumptions made initially, this method avoids the trap of simplification through imposing {\em a priori} assumptions merely by {\em fiat}. Another recent effort \cite{CPFSGBL-CPL2010} in the same spirit considered multi-qubit states that are well represented by matrix product states \cite{KSZ1991,FNW1992,P-GVWC2007} (which require a number of parameters growing only polynomially with the number of qubits). Many states of interest, such as ground states of certain model Hamiltonians in condensed-matter physics, are of that form.
Crucially, the particular form of the state can be verified by the data.

Here we go one step further, and we will allow, tentatively, {\em any} parametrized form for the density matrix of the quantum system to be tested, possibly containing just a few  parameters. In fact, we may have several different tentative ideas of how our quantum state is best parameterized. The questions are then, how the data reveal which of those descriptions work sufficiently well, and which description is the best.
This  idea corresponds to a well-developed field in statistics: model selection \cite{J1961b,BA2002b,Z2000}. All mathematical descriptions of reality are in fact models (and a quantum state, pure or mixed, is an excellent example of a model), and they can be evaluated by judging their performance relative to that of the true model (assuming it exists). In order to quantify this relative performance, we will make use of the Kullback-Leibler divergence (aka mutual information, aka cross entropy, aka relative entropy) \cite{KL1951}, which has the interpretation of the amount of information lost when a specific model is used rather than the true model.
Based on the minimization of the Kullback-Leibler divergence over different models, the Akaike Information Criterion (AIC) \cite{A1974} was developed as a ranking system so that models are evaluated with respect to each other, given measurement data. The only quantities appearing in the criterion are the maximum likelihood obtainable with a given model (i.e., the probability the observed data would occur according to the model, maximized over all model parameters), and the number of independent parameters of the model.

The minimization does not require any knowledge of the true model, only that the testing model is sufficiently close to the true model. The legitimate application of AIC should, therefore, in principle be limited to ``good'' models, ones that include the true model (in our case, the exact quantum state that generated the data), at least to a very good approximation. However this does not prevent one from resorting to the AIC for model evaluation when there is no such guarantee. In fact, Takeuchi studied the case where the true model does not belong to the model set and came up with a more general criterion, named the Takeuchi Information Criterion, TIC \cite{T1976}. However the estimation of the term introduced by Takeuchi to counterbalance the bias of the maximum likelihood estimator used in the AIC, requires estimation of a $K\times K$ matrix ($K$ being the number of independent parameters used by a model) from the data, which, unfortunately, is prone to significant error. This reduces the overall charm and practical use of the TIC. Since in most cases the AIC is still a good approximation to the TIC \cite{BA2002b}, especially in the case of many data, we stick to the simpler and more robust criterion here.

Information criteria are designed to produce a relative (rather than absolute) ranking of models, so that fixing a reference model is  convenient. Throughout this paper we choose the ``full-parameter model'' (FPM) as reference, that is, a model with just enough independent variables to fully parameterize the measurement on our quantum system. For tomographically complete measurements (discussed in detail in Sec.~\ref{sec_tomography}) the number of independent variables is given by the number of free parameters in the density matrix ($2^{2D}-1$ for a $D$-dimensional Hilbert space). For tomographically incomplete measurements (see Sec.~\ref{sec_witness}), the number of independent variables of FPM is smaller, and equals the number of independent observables.  We will, in fact, not even need the explicit form of the FPM (which may be hard to construct for tomographically incomplete measurements), as its maximum possible likelihood can be easily upper-bounded.

We should note an important distinction between maximum likelihood estimation (MLE) \cite{BDPS1999}, a technique often used in quantum tomography, and the method of information criteria and model selection. MLE produces the state that fits the data best. Now the data inevitably contains (statistical) noise, and the MLE state predicts, incorrectly, that same noise to appear in future data. Information criteria, on the other hand, have been designed to find the model that best predicts future data, and tries, in particular, to avoid overfitting to  the data, by limiting the number of model parameters. This is how a model with a few parameters can turn out to be the best predictive model, even if, obviously, the MLE state will fit the (past) data better.

We also note that information criteria have been applied mostly in areas of research outside of physics. This is simply due to the happy circumstance that in physics we tend to know what the ``true'' model underlying our observations is (or should be), whereas this is much less the case in other fields. Within physics, information criteria have been applied to astrophysics \cite{L2007}, where one indeed may not know the ``true'' model (yet), but also to the problem of entanglement estimation \cite{LvE2009}. In the latter case (and in quantum information theory in general) the problem is not that we do not know what the underlying model is, but that that model may contain far too many parameters. Hence the potential usefulness of information criteria. And as we recently discovered, the AIC has even been applied to quantum state estimation, not for the purpose of making it more efficient, but making it more accurate, by avoiding overfitting \cite{UNTMN2003}.

\section{The Akaike Information Criterion - A Schematic Derivation}
\label{sec_AIC}

Suppose we are interested in measuring certain variables, summarized as a vector $\mathbf{x}$, and their probability of occurrence as outcome of our measurement. We denote $f(\mathbf{x})$ as the probabilistic model that truthfully reflects reality (assuming for convenience that such a model exists) and $g(\mathbf{x}|\vec{\theta})$ as our (approximate) model characterized by one or more parameters, summarized as a vector $\vec{\theta}$. The models satisfy the normalization condition $\int {\rm d}\mathbf{x}f(\mathbf{x})=\int{\rm d}\mathbf{x}g(\mathbf{x}|\vec{\theta})=1$ for all $\vec{\theta}$. By definition, we say there is no information lost when $f(\mathbf{x})$ is used to describe reality. The amount of information lost when $g(\mathbf{x}|\vec{\theta})$ is used instead of the true model is defined to be the Kullback-Leibler divergence \cite{KL1951} between the model $g(\mathbf{x}|\vec{\theta})$ and the true model $f(\mathbf{x})$:
\begin{align}
  I(f,g_{\vec{\theta}})=&\int {\rm d}\mathbf{x} f(\mathbf{x})\log(f(\mathbf{x}))\nonumber\\
  &-\int {\rm d}\mathbf{x} f(\mathbf{x})\log(g(\mathbf{x}|\vec{\theta})).
  \label{eq_Ifg_int}
\end{align}
Eq.~(\ref{eq_Ifg_int}) can be conveniently rewritten as
\begin{align}
  I(f,g_{\vec{\theta}})=E_\mathbf{x}\left[\log(f(\mathbf{x}))\right]-E_\mathbf{x}\left[\log(g(\mathbf{x}|\vec{\theta}))\right],
  \label{eq_Ifg_exp}
\end{align}
where $E_\mathbf{x}[\cdot]$ denotes an estimate with respect to the true distribution $f(\mathbf{x})$.
 We see that $\mathbf{x}$ is no longer a variable in the above estimator, as we integrated it out. The only variable that affects $I(f,g_{\vec{\theta}})$ is $\vec{\theta}$. Since the first term in Eq.~(\ref{eq_Ifg_int}) is irrelevant to the purpose of rank-ordering different models $g$ (not to mention we cannot evaluate it when $f$ is not known), we only have to consider the second term. Suppose there exists $\vec{\theta}_0$ such that $g(\mathbf{x}|\vec{\theta}_0)=f(\mathbf{x})$ for every $\mathbf{x}$, that is, the true model is included in the model set. Note that for this to hold, $\vec{\theta}$ does not necessarily contain the same number of parameters as the dimension of the system. To use a simpler notation without the integration over $\mathbf{x}$ we denote the second term in Eq.~(\ref{eq_Ifg_int}) (without the minus sign) as
\begin{align}
  S(\vec{\theta}_0:\vec{\theta})=\int {\rm d}\mathbf{x} g(\mathbf{x}|\vec{\theta}_0)\log(g(\mathbf{x}|\vec{\theta})),
  \label{eq_S_int}
\end{align}
where we have used $g(\mathbf{x}|\vec{\theta}_0)$ to represent the true model $f(\mathbf{x})$. The advantage of this estimator is that it can be approximated without knowing the true distribution $f(\mathbf{x})$. To do that we first consider the situation where $\vec{\theta}$ is close to $\vec{\theta}_0$. This assumption can be justified in the limit of large $N$, $N$ being the number of measurement records, since the model $\vec{\theta}$ ought to approach $\vec{\theta}_0$ asymptotically (assuming, for simplicity, $\vec{\theta}_0$ is unique). We know that $S(\vec{\theta}_0:\vec{\theta})$ must have a maximum when $\vec{\theta}=\vec{\theta}_0$, and we may then symbolically expand $S(\vec{\theta}_0:\vec{\theta})$ in the vicinity of $\vec{\theta}_0$ by
\begin{align}
  S(\vec{\theta}_0:\vec{\theta})= S(\vec{\theta}_0:\vec{\theta}_0)&-\frac{1}{2}||\vec{\theta}-\vec{\theta}_0||_{\vec{\theta}_0}^2\nonumber\\
  &+{\cal O}\left(||\vec{\theta}-\vec{\theta}_0||_{\vec{\theta}_0}^{3/2}\right),
  \label{eq_S_theta0-theta}
\end{align}
where
\begin{align}
  ||\vec{\theta}-\vec{\theta}_0||_{\vec{\theta}_0}^2=\left(\vec{\theta}-\vec{\theta}_0\right)'\cdot\left.\frac{\partial^2S(\vec{\theta}_0:\vec{\theta})}{\partial\vec{\theta}^2}\right|_{\vec{\theta}=\vec{\theta}_0}\cdot\left(\vec{\theta}-\vec{\theta}_0\right)
\end{align}
denoting a \emph{squared length} derived from a metric defined at $\vec{\theta}_0$. It can be proved that when $N$ is sufficiently large
$||\vec{\theta}-\vec{\theta}_0||_{\vec{\theta}_0}^2$ can be approximated by the $\chi_K^2$ distribution, with $K$ equal to the number of independent parameters used by the model $\vec{\theta}$. From the properties of the $\chi_K^2$ distribution, we know the average value of $||\vec{\theta}-\vec{\theta}_0||_{\vec{\theta}_0}^2$ will approach $K$.

The next step is to evaluate the estimator $S(\vec{\theta}_0:\vec{\theta}_0)$, where $\vec{\theta}_0$ is now considered a variable. Suppose we find the maximum likelihood estimate $\vec{\theta}_M$ from the measurement outcomes such that $S(\vec{\theta}_M:\vec{\theta}_M)$ is the maximum. Now $\vec{\theta}_M$ should also be close to the true model $\vec{\theta}_0$, when $N$ is sufficiently large. Therefore we can similarly expand $S(\vec{\theta}_0:\vec{\theta}_0)$ in the vicinity of $\vec{\theta}_M$ as
\begin{align}
  S(\vec{\theta}_0:\vec{\theta}_0)= S(\vec{\theta}_M:\vec{\theta}_M)&-\frac{1}{2}||\vec{\theta}_0-\vec{\theta}_M||_{\vec{\theta}_M}^2\nonumber\\
  &+{\cal O}\left(||\vec{\theta}-\vec{\theta}_0||_{\vec{\theta}_M}^{3/2}\right),
  \label{eq_S_theta0-theta0}
\end{align}
$||.||_{\vec{\theta}_M}$ is a length similarly defined as in Eq.~(\ref{eq_S_theta0-theta}) and has the same statistical attributes
as $||\vec{\theta}-\vec{\theta}_0||_{\vec{\theta}_0}^2$ since $\vec{\theta}_0$ is related to $\vec{\theta}_M$ the same way $\vec{\theta}$ is related to $\vec{\theta}_0$ and $\vec{\theta}_M$ is very close to $\vec{\theta}_0$. Its average value, therefore, approaches again $K$, according to the $\chi_K^2$ distribution. Thus we are able to rewrite Eq.~(\ref{eq_S_int}) as
\begin{align}
S(\vec{\theta}_0:\vec{\theta})\approx S(\vec{\theta}_M:\vec{\theta}_M)-K.
\end{align}
We see that now our target estimator $S(\vec{\theta}_0:\vec{\theta})$ is evaluated by the MLE solution $\vec{\theta}_M$ only (plus the number of parameters $K$ of the model), with no knowledge of what the true model $f$ is. The assumption that underlies this convenience is constituted by two parts: estimating $S(\vec{\theta}_0:\vec{\theta})$ with its maximum $\vec{\theta}_0$ and estimating $S(\vec{\theta_0}:\vec{\theta}_0)$ from the data by its optimum $\vec{\theta}_M$. The deviations from their respective maxima are equal and result simply in the appearance of the constant $K$.

We now denote $\L_M=S(\vec{\theta}_M:\vec{\theta}_M)$, which is the maximum likelihood obtainable by our model, with respect to a given set of measurement records. The AIC is then defined by
\begin{align}
{\rm AIC}=-2\L_M+2K.\label{eq_AIC}
\end{align}
Apart from the conventional factor 2,  and a constant independent of the model $\vec{\theta}$, AIC is an estimator of the quantity in Eq.~(\ref{eq_Ifg_int}) we originally considered, that is, the Kullback-Leibler divergence between a model that is used to describe the true model and the true model itself. Therefore a given model is considered better than another if it has a {\em lower} value of AIC.

Finally, in the case that $N$ is not so large yet that asymptotic relations hold to a very good approximation, one
can include a correction factor to the AIC taking the deviation from asymptotic values into account. The corrected AIC gives rise to a slightly different criterion \cite{Sug1978}:
\begin{align}
{\rm AICc}=-2\L_M+2K+\frac{2K(K+1)}{N-K-1}.\label{eq_AICc}
\end{align}

\section{Results}
\label{sec_results}

\subsection{Dicke states}
\label{sec_Dicke}

We will apply the AIC to measurements on a popular family of entangled states, the Dicke states of four qubits \cite{D1954,HHRBCCKRRSBGDB2005,KSTSW2007,PCDLvEK2009,CGPvEK2010}. We simulate two different experiments, one tomographically complete experiment, another measuring an entanglement witness. We include imperfections of a simple type, and we investigate how model selection, according to the AIC, would work. We consider cases where we happen to guess the correct model, as well as cases where our initial guess is, in fact, incorrect.

We consider the four-qubit Dicke states with one or two excitations $\ket{D_4^{1,2}}$ (with the state $\ket{1}$ representing an excitation):
\begin{subequations}
\begin{align}
  \ket{D_4^1}=&\left(\ket{0001}+\ket{0010}+\ket{0100}+\ket{1000}\right)/2,\\
  \ket{D_4^2}=&\left(\ket{0011}+\ket{0101}+\ket{0110}+\ket{1001}\right.\nonumber\\
  &+\left.\ket{1010}+\ket{1100}\right)\sqrt{6}.
\end{align}
\end{subequations}
For simplicity, let us suppose that white noise is the only random noise in the state generation,
and that it corresponds to mixing of the ideal state with the maximally mixed state of the entire space (instead of the subspace with exactly one or two excitations, which could be a reasonable choice, too, depending on the actual implementation of the Dicke states). We thus write the states under discussion as
\begin{align}
  \rho^{1,2}(\alpha)=(1-\alpha)\ket{D_4^{1,2}}\bra{D_4^{1,2}}+\alpha\openone/D,
\end{align}
where $\openone/D$ is the maximally mixed state for dimension $D=2^4$, and $0\leq \alpha\leq 1$. We will fix the actual state generating our data to be
\begin{align}
  \rho_{\rm actual}^{1,2}=\rho^{1,2}(\alpha=0.2).
\end{align}
This choice is such that the mixed state is entangled (as measured by our multi-qubit version of the negativity, see below), even though the entanglement witness whose measurement we consider later in Sec.~\ref{sec_witness}, just fails to detect it.

For our first model (to be tested by AIC) we wish to pick a one-parameter model (so, $K=1$) that also includes a wrong guess.
A straightforward model choice, denoted by $\mone$, is
\begin{align}
  \mone: \rho_\phi^{1,2}(q)=(1-q)\ket{\Psi_{\rm target}^{1,2}(\phi)}&\nonumber\\
 \bra{\Psi_{\rm target}^{1,2}(\phi)}+&q\openone/D.
  \label{eq_M1_SICPOVM}
\end{align}
We refer to the pure states appearing here as the {\em target} states $\ket{\Psi_{\rm target}^{1,2}(\phi)}$, simulating the case where we (possibly incorrectly) think we would be creating a pure state of that form, if only the white noise were absent ($q=0$). The phase $\phi$ is included not as a (variable) parameter of the model but as an inadvertently mis-specified property. In this case, it stands for us being wrong about a single relative phase in one of the qubits in state $|1\rangle$.
Without loss of generality we assume the first qubit in our representation to carry the wrong phase, and we write
\begin{subequations}
\begin{align}
  \ket{\Psi_{\rm target}^1(\phi)}=&\frac{1}{2}\left(\ket{0001}+\ket{0010}+\ket{0100}\right.\nonumber\\
  &+\left.e^{i\phi}\ket{1000}\right),\label{eq_tilde1}\\
  \ket{\Psi_{\rm target}^2(\phi)}=&\frac{1}{\sqrt{6}}\left[\ket{0011}+\ket{0100}+\ket{0110}\right.\nonumber\\
  +&\left.e^{i\phi}\left(\ket{1001}+\ket{1010}+\ket{1100}\right)\right].\label{eq_tilde2}
\end{align}
\label{eq_tilde}
\end{subequations}
Alternatively, if we {\em do} consider this a two-parameter model (changing $K=1$ to $K=2$), then $\phi$ is variable, and we would optimize over $\phi$. In our case, this optimum value should always be close to $\phi=0$.

\begin{figure}[t]
  \begin{center}
    \subfigure[$\mone$.]
    {\includegraphics[width=195pt]{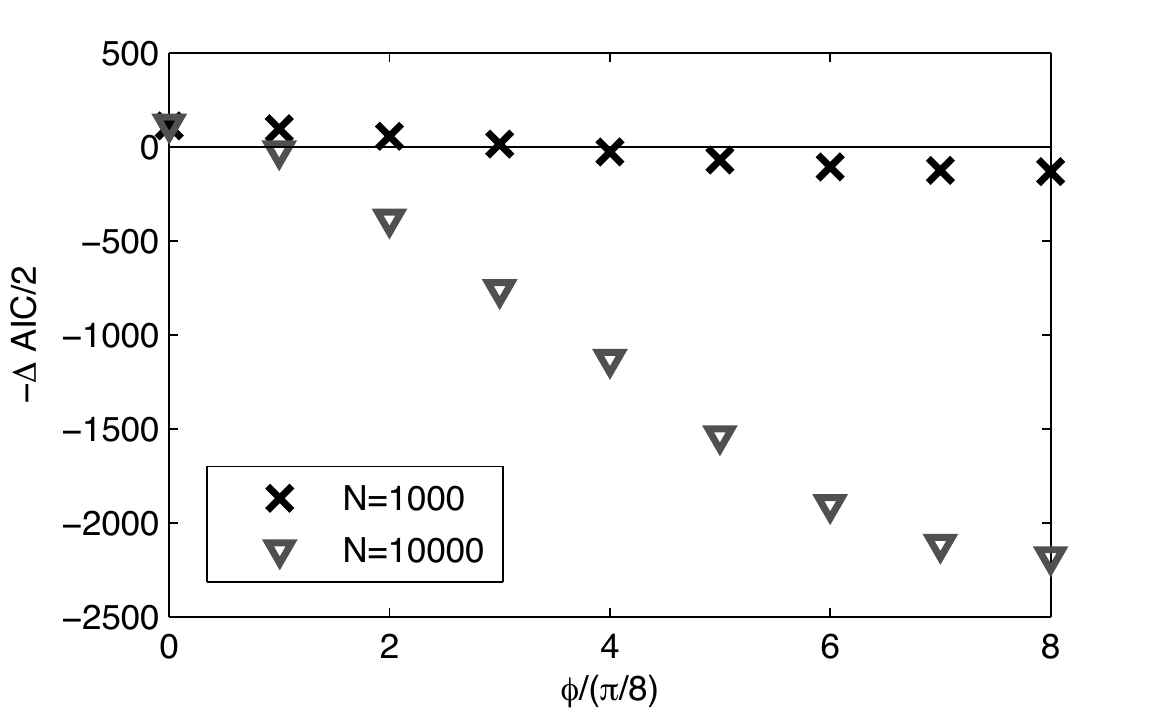}}
    \\
    \subfigure[$\mtwo$.]
    {\includegraphics[width=195pt]{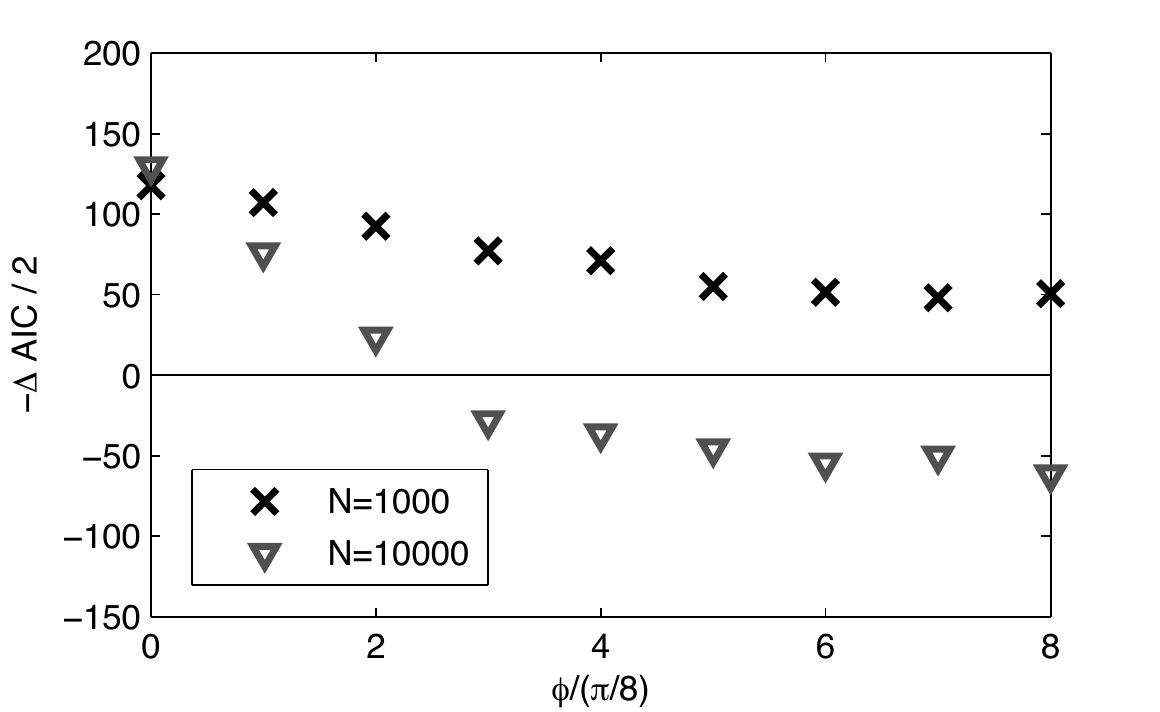}}
  \end{center}
  \caption{{\em How AIC ranks the one- and two-parameter models vs. the full-parameter model (FPM):} \\
    Plot of the difference between AIC values of our models and the FPM, \textsl{i.e.}, $-\Delta{\rm AIC}={\rm AIC(FPM)}-{\rm AIC}(\mone)$ or $-\Delta{\rm AIC}={\rm AIC(FPM)}-{\rm AIC}(\mtwo)$, for various numbers of SIC-POVM measurements, $N$, with $\ket{\Psi_{\rm target}^1}$ as the target state, as  functions of the angle $\phi$. The horizontal line demarcates $\Delta{\rm AIC}=0$: points above (below) that line correspond to cases where the model with fewer (more) parameters is preferred. The figures with $\ket{\Psi_{\rm target}^2}$ as the target state look very similar (see FIG.~\ref{fig_AIC_M1_1e4_SICPOVM} for an example of this similarity).}
  \label{fig_AIC_M1M2_SICPOVM}
\end{figure}

\subsection{Tomographically complete measurement}
\label{sec_tomography}
We first consider a tomographically complete measurement, in which a so-called SIC-POVM (symmetric informationally complete positive operator values measure  \cite{RK-BSC2004}) with 4 outcomes is applied to each qubit individually. We first test our one-parameter model, and compare it to the FPM, which  contains 255 ($=4^4-1$) parameters, which is the number of parameters needed to fully describe a general state of 4 qubits. With definition Eq.~(\ref{eq_AIC}) we have
\begin{align}
  {\rm AIC}(\mone)=-2\L_M(\mone)+2,
  \label{eq_AIC_M1_SICPOVM}
\end{align}
since $K=1$ for $\mone$. For the FPM we have
\begin{align}
  {\rm AIC(FPM)} =-2\L_M({\rm FPM})+2\times 255,
  \label{eq_AIC_FPM_SICPOVM}
\end{align}
where $\L_M({\rm FPM})$ is the log of the maximum likelihood obtainable by the FPM. The latter can be bounded from above by noting that the best possible FPM would generate probabilities that exactly match the actual observed frequencies of all measurement outcomes. In the following we will always use that upper bound, rather than the actual maximum likelihood. Even though it is possible to find the maximum likelihood state in principle (and even in practice for small enough Hilbert spaces), we are only concerned with the FPM's ranking according to the AIC, which does not require its density matrix representation. For $\mone$ to beat the FPM we require
\begin{align}
  -\Delta{\rm AIC}:={\rm AIC}({\rm FPM})-{\rm AIC}(\mone)>0.
\end{align}
This is a sufficient but not necessary requirement, as we use the above-mentioned upper bound to the FPM likelihood.

\begin{figure}[t]
  \begin{center}
    \includegraphics[width=195pt]{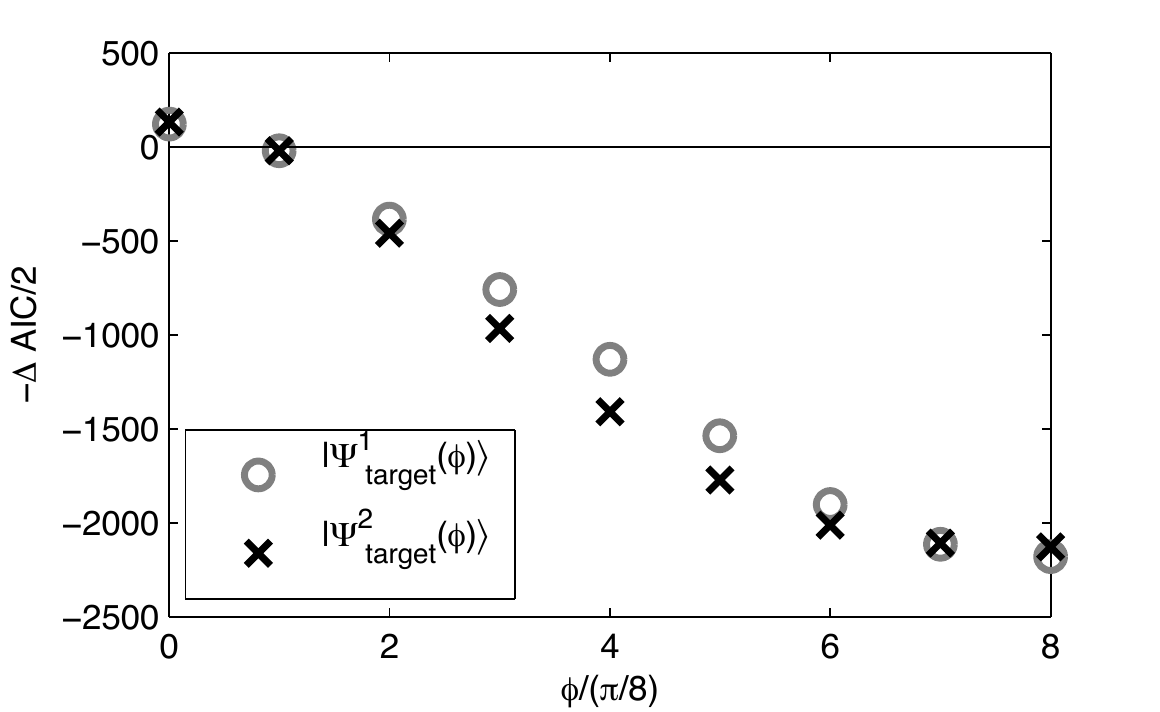}
  \end{center}
  \caption{{\em Comparing single- and double-excitation Dicke states}: The difference between AICs of $\mone$ and the FPM, \textsl{i.e.}, $-\Delta{\rm AIC}={\rm AIC(FPM)}-{\rm AIC}(\mone)$ for both target states, when $N=10000$, as functions of $\phi$. The horizontal line demarcates $\Delta{\rm AIC}=0$. }
  \label{fig_AIC_M1_1e4_SICPOVM}
\end{figure}

We plot the difference $\Delta{\rm AIC}$ between the two rankings in FIG.~\ref{fig_AIC_M1M2_SICPOVM}(a) for various values of the number of measurements, and for various values of the phase $\phi$. We observe the following: The simple model
is, correctly, judged better than the FPM when the phase $\phi$ is sufficiently small.
The more measurements one performs, the smaller $\phi$ has to be for AIC to still declare the model superior to the FPM (i.e., for the points to stay above the solid line, at $\Delta{\rm AIC}=0$).

Although the correction to the AIC mentioned in Eq.~(\ref{eq_AICc}) is not very small for the FPM for $N=1000$, applying that correction still does not shift the second and third point below zero: that is, $N=1000$ measurements is still not sufficiently large for the AICc to recognize that $\phi=\pi/4$ and $\phi=\pi/2$ are incorrect guesses. One can argue about what the cause of this is: it could be that $N$ is just too small for the derivation of the AIC (or even the AICc) to be correct. Or it could be that the AIC ranking is unreliable because the assumption that the true model is included in the model, is violated. Or it could be that, even with a perfectly valid criterion (perhaps the TIC), the statistical noise present in the data would still be too large.

If we consider the phase $\phi$ as a second (variable) parameter (thus creating a two-parameter model), then we can give FIG.~1 a different interpretation: we would pick $\phi=0$ as the best choice, and we would increase $K$ by 1. The latter correction is small on the scale of the plots, and so we find the two-parameter model to be superior to the one-parameter model for any nonzero plotted value of $\phi$, and to the FPM.
This is a good illustration of
the following rather obvious fact:
even if one has the impression that a particular property of one's quantum source is (or ought to be) known, it still might pay off to represent that property explicitly as a variable parameter (at the small cost of increasing $K$ by 1), and let the data determine its best value.

\subsection{Cross modeling}
Suppose one picked a one-parameter model with a wrong (nonzero) value of $\phi$, and the AIC has declared the model to be worse than the FPM. How can one improve the model in a systematic way when one lacks a good idea of which parameters to add to the model (we assume we already incorporated all parameters deemed important {\em a priori}). Apart from taking more and different measurements, one could use a hint from the existing data.
One method making use of the data is to apply ``cross modeling,'' where half the data is used to construct a modification to the model, and the remaining half is used for model validation, again by evaluating AIC on just that part of the data. So suppose $N$ measurements generate a data sequence $\mathcal{F}=\{f_1,f_2,...,f_N\}$. One takes, \textsl{e.g.}, the first $N/2$ data points, $\{f_1,...,f_{N/2}\}$, as the training set, and acquires the MLE state $\rho_{\rm MLE}$, or a numerically feasible approximation thereof,  with respect to the training set. We then create a model with two parameters like so:
\begin{align}
  \mtwo:\rho_\phi(\epsilon,q)=(1-\epsilon)\left[(1-q)\rho_{\rm MLE}\right.&\nonumber\\
  \left.+q\ket{\Psi_{\rm target}^{1,2}(\phi)}\bra{\Psi_{\rm target}^{1,2}(\phi)}\right]+\epsilon\openone/D.&
  \label{eq_M2_SICPOVM}
\end{align}
For practical reasons $\rho_{\rm MLE}$ does not need to be strictly the MLE state, in particular when the dimension of the full parameter space is large. One would only require it to explain $\{f_1,...,f_{N/2}\}$ well enough to make sure that part of the data is properly incorporated in the model. Thus, one could, for example, use one of the numerical shortcuts described in \cite{KJ2009}.  The rest of the data $\{f_{N/2+1},...,f_N\}$ is used to evaluate $\mtwo$ against the FPM.

We note the resemblance of this procedure with the method of ``cross-validation'' \cite{SEJS1986}. In cross-validation one tries to find out how well a {\em given} predictive model performs by partitioning the data set into training set and validation set (exactly the same idea as given above). One uses multiple different partitions, and the results are averaged and optimized over those partitions. It can be shown \cite{S1977} that under certain conditions cross-validation and the AIC are asymptotically equivalent in model selection. This virtually exempts one from having to check multiple partitions of the data set, by applying the AIC to the whole data set.

It is worth emphasizing that what we do here is different in two ways. First, our model is not fixed but modified, based on information obtained from one half of the data. Second, we  partition the data set only once, and the reason is, that it would be cheating to calculate the (approximate) MLE state of the full set of data (or, similarly, check many partitions and average), and then consider the resulting MLE state a parameter-free model.

FIG.~\ref{fig_AIC_M1M2_SICPOVM}(b) shows results for $\mtwo$ and SIC-POVM measurements. When the number of measurements is $N=1000$, all $\mtwo$ models
are considered better than the FPM,  regardless of the phase error $\phi$ assumed for the target state. The reason is that around $\phi=\pi/2$ the approximate MLE state obtained from the first half of the data is able to ``predict'' the measurement outcomes (including their large amount of noise!) on the second half better than the 1-parameter model with the wrong phase.

On the other hand, when $N=10000$ the AIC recognizes only the simple models with small phase errors ($\phi=0, \pi/8, \pi/4$) as better than the FPM.
So, neither the approximate MLE state, nor the 1-parameter model with wrong phase are performing well. This indicates how many measurements are needed to predict a single phase to a given precision.

\begin{figure}[t]
  \begin{center}
      \includegraphics[width=195pt]{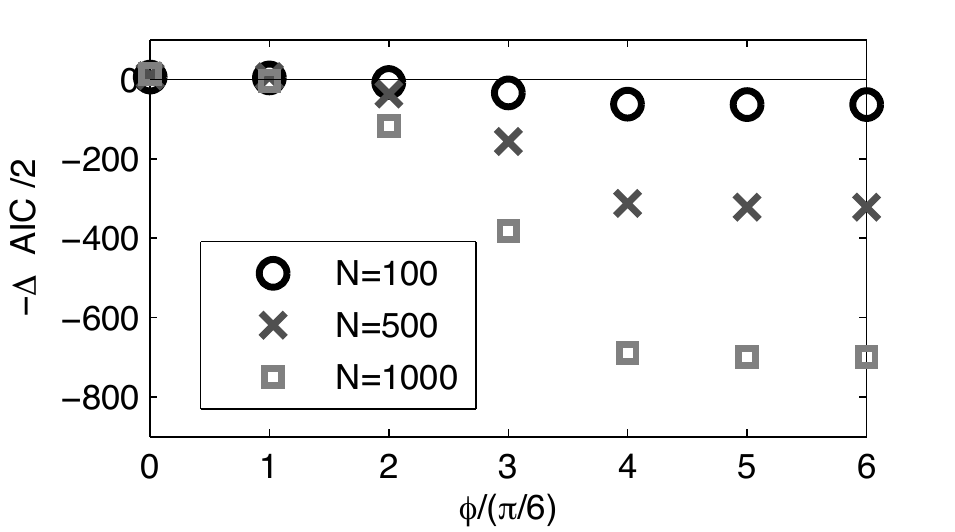}
  \end{center}
  \caption{{\em How the AIC ranks our one-parameter model vs. the FPM for an entanglement witness measurement}:
  The difference between AICs of $\mone$ with $\ket{\Psi_{\rm target}^2(\phi)}$ and the FPM, \textsl{i.e.}, $-\Delta{\rm AIC}={\rm AIC(FPM)}-{\rm AIC}(\mone)$, for different numbers of witness measurements, as functions of $\phi$. The horizontal line demarcates $\Delta{\rm AIC}=0$.}
  \label{fig_AIC_M1_witness}
\end{figure}

\subsection{Witness measurement}
\label{sec_witness}

For states that are close to symmetric Dicke states $\ket{D_N^{N/2}}$, their entanglement can be verified by using measurements that require only two different local settings, \textsl{e.g.}, spins (or polarizations) either {\em all} in the $x$-direction or {\em all} in the $y$-direction. In particular, when $N=4$, an efficient witness is $W_{Jxy}=7/2+\sqrt{3}-J_x^2-J_y^2$ \cite{T2007}, where $J_{x,y}=\sum_j\sigma_{x,y}^{(j)}/2$, with $\sigma_{x,y}^{(j)}$ the Pauli matrices for the $j$-th subsystem. This witness detects (by having a negative expectation value) Dicke states  with a white noise background, \textsl{i.e.}, $\rho(\alpha)=(1-\alpha)\ket{D_4^2}\bra{D_4^2}+\alpha\openone/D$ whenever $0\leq\alpha<0.1920$.

\begin{figure}[t]
\begin{center}
  \includegraphics[width=195pt]{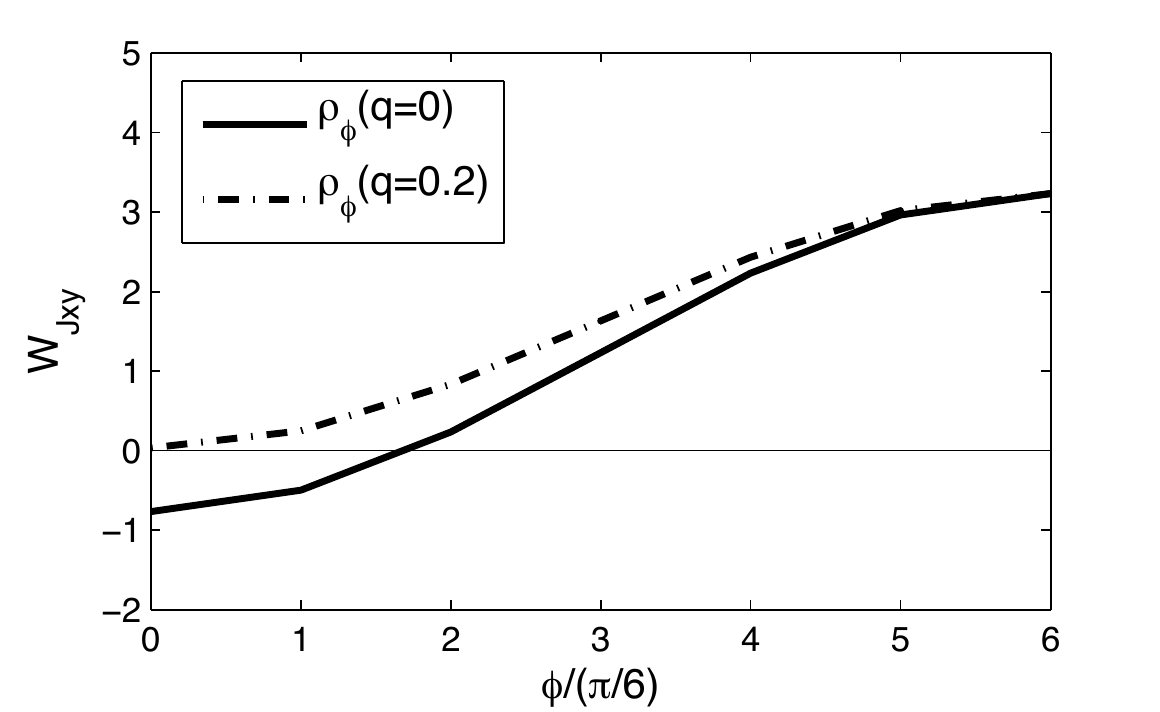}
  \caption{{\em Does the witness $W_{J_{xy}}$ detect entanglement if there is a phase error?}:  Witness performance $\langle W_{J_{xy}}\rangle$ for different states (defined as in Eq.~(\ref{eq_M1_witness})) as a function of $\phi$. A negative expectation value detects entanglement.}
  \label{fig_WJxy}
\end{center}
\end{figure}

So we suppose we perform $N/2$ measurements on all of the four spins in the $x$-direction simultaneously, and another $N/2$ similar measurements in the $y$-direction. Instead of calculating the witness $W_{J_{xy}}$ and ending up with one single value determining entanglement, we make use of the full record of all individual outcomes in order to evaluate (and then maximize)  likelihoods. For example, for the measurement of all four spins in the $x$-direction simultaneously, we can count the number of times they are projected onto the $\ket{x+x+x+x+}$ state, the $\ket{x+x+x+x-}$ state, etc. In both $x$- or $y$-directions, the number of independent observables (i.e., the number of independent joint expectation values) is 15, which can be seen as follows: Any density matrix of $M$ qubits can be expressed in terms of the expectation values of $4^M$ tensor products of the 3 Pauli operators and the identity $\openone$, but the expectation value of the product of $M$ identities equals 1 for any density matrix, thus leading to $4^M-1$ independent parameters encoded in a general density matrix. From having measured just  $\sigma_x$ on all $M$ qubits, we can evaluate all expectation values of all operators that are tensor products of $\sigma_x$ and the identity. There are $2^M$ such products, and subtracting the trivial expectation value for $\openone^{\otimes M}$ leaves $2^M-1$ independent expectation values.

This means it only takes $2\times 15=30$ independent parameters to form the FPM, and we have $K=30$. Similar to the  tomographically complete case, we do not need the concrete form of the whole 255-30 dimensional manifold of MLE states, nor do we need to explicitly parameterize the 30-parameter FPM states, as we can simply upper bound the maximum likelihood for this model, $L_M({\rm FPM})$, by noting the best one could possibly do is reproduce exactly the observed frequencies of all possible measurement outcomes.

\begin{figure}[t]
\begin{center}
  \subfigure[$N=100$]{\includegraphics[width=195pt]{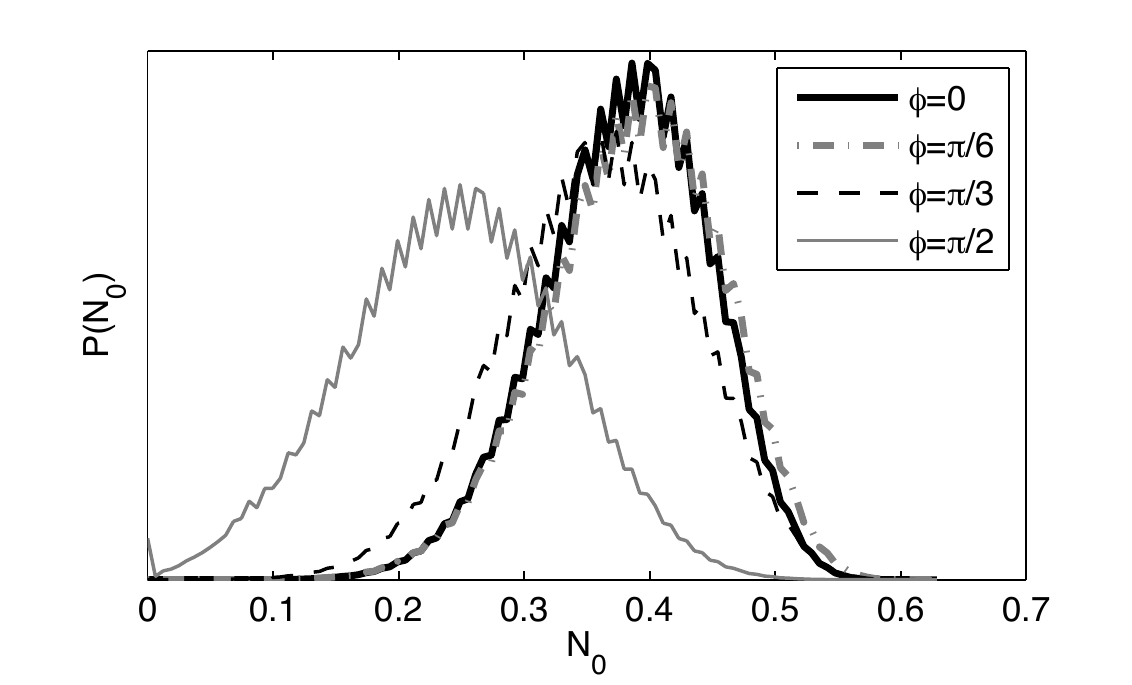}}
  \subfigure[$N=1000$]{\includegraphics[width=195pt]{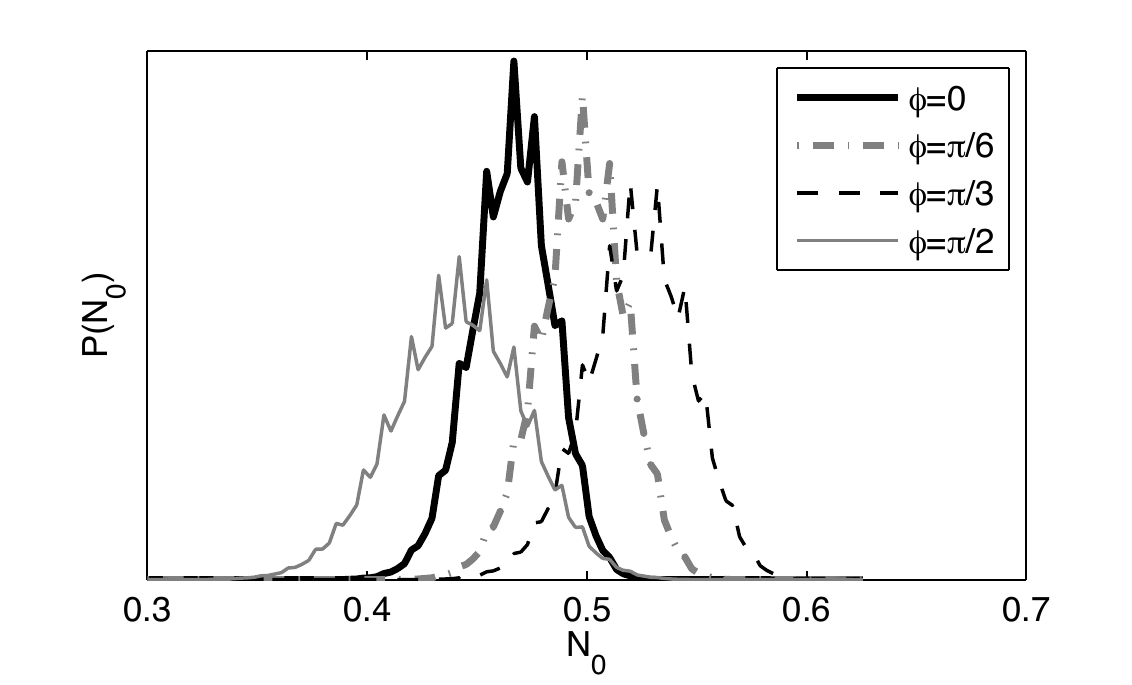}}
  \caption{{\em How one quantifies entanglement from a witness measurement}:
  The posterior probability distributions of $\mathcal{N}_0$ for different numbers of witness measurements from model $M_1(q)$ with target state $\ket{\Psi_{\rm target}^{2(1)}}$, where $\phi=0, \pi/6, \pi/3$. $\mathcal{N}_0(\rho_{\rm actual})=0.4770$. The prior distribution is assumed to be uniform  on [0,1] for both $\epsilon$ and $q$. The distributions of $\mathcal{N}_1$ and $\mathcal{N}_2$ are similar (up to a simple shift).}
  \label{fig_negdist_M1_witness}
\end{center}
\end{figure}

\subsection{Estimating entanglement}
Our state $\rho_{\rm actual}=\rho(\alpha=0.2)$ is just not detected by the witness $W_{Jxy}$, but still contains a considerable amount of entanglement.
We choose to quantify this entanglement by means of three entanglement monotones (of which only two are independent), simply constructed from all bipartite negativities.
If the four parties are denoted $A$, $B$, $C$ and $D$, the generalized negativities \cite{VW2002,SG-A2008,YvE2011} are defined as
\begin{subequations}
\begin{align}
  \mathcal{N}_1=&\left(\mathcal{N}_{AB-CD}\mathcal{N}_{AC-BD}\mathcal{N}_{AD-BC}\right)^{1/3},\\
  \mathcal{N}_2=&\left(\mathcal{N}_{A-BCD}\mathcal{N}_{B-CDA}\mathcal{N}_{C-DAB}
  \mathcal{N}_{D-ABC}\right)^{1/4},\\
  \mathcal{N}_0=&\left(\mathcal{N}_1^3 \mathcal{N}_2^4\right)^{1/7},
\end{align}
\end{subequations}
where $\mathcal{N}_{AB-CD}$ denotes the negativity with respect to partition $AB$ against $CD$, etc. The main advantage of the generalized negativities is that they are all efficiently computable directly from the density matrix.
We have for our state
$ \mathcal{N}_1=0.6293$, $ \mathcal{N}_2=0.3875$, and $ \mathcal{N}_0=0.4770$.

Similarly to the tomographically complete case, we first consider the following one-parameter model:
\begin{align}
  \mone:~\rho_\phi(q)=(1-q)\ket{\Psi_{\rm target}^2(\phi)}&\nonumber\\
  \bra{\Psi_{\rm target}^2(\phi)}+q&\openone/D,\label{eq_M1_witness}
\end{align}
where $\ket{\Psi_{\rm target}^2(\phi)}$ is defined in Eg.~(\ref{eq_tilde1}). The AICs for $\mone$ and FPM are
\begin{align}
  {\rm AIC}(\mone)=&-2\L_M(\mone)+2,\label{eq_AIC_M1_witness}\\
  {\rm AIC}({\rm FPM})=&-2\L_M({\rm FPM})+2\times30.\label{eq_AIC_FPM_witness}
\end{align}
FIG.~\ref{fig_AIC_M1_witness} shows that, as before, the marks above the horizontal solid line correspond to models deemed better than FPM. Compared to the case of full tomography (FIG.~\ref{fig_AIC_M1M2_SICPOVM}(a)), here  the value of ${\rm AIC}(\mone)$ is larger than ${\rm AIC}({\rm FPM})$ by a much smaller amount, even when the phase term is correct ($\phi=0$). The absolute value of the difference is not  relevant, though, and what counts is its sign. The obvious reason for the smaller difference is that the number of independent parameters for the FPM has dropped from 255 to 30. In addition, the FPM in this case does not refer to a specific 30-parameter model. On the contrary, since the number of degrees of freedom of the quantum system is still 255, there is a whole subspace of states, spanning  a number of degrees of freedom equal to 225 (=255-30), all satisfying the maximum likelihood condition.

The witness measurement is very sensitive to the phase error, even when the number of measurements is still small. When $N=1000$, the estimation of $\phi$ is within an error of $\pi/6$, as the second point plotted is already below the line $\Delta{\rm AIC}=0$.
Compared to FIG.~\ref{fig_AIC_M1M2_SICPOVM}(a), this precision is only reached when $N=10000$.

An interesting comparison can be made between AIC and the entanglement-detecting nature of witness $W_{J_{xy}}$. FIG.~\ref{fig_WJxy} shows the performance of $\langle W_{J_{xy}}\rangle$ for the pure state $\ket{\Psi_{\rm target}^2(\phi)}$ ($\rho_\phi(q=0)$, solid curve) and the mixed with 20\% of identity mixed in ($\rho_\phi(q=0.2)$, dot-dashed curve). Even when the state is pure, $\langle W_{J_{xy}}\rangle$ will not be able to witness any entanglement if the phase error is larger than $\pi/3$, just about when AIC declares such a model deficient. Entanglement in the mixed $\rho_\phi(q=0.2)$ of course is never witnessed. This means $\langle W_{J_{xy}}\rangle$ is only an effective witness in the vicinity of $\ket{D_4^2}$, with limited tolerance of either white noise or phase noise in even just one of the four qubits. (Of course, one would detect the entanglement in the pure state by appropriately rotating the axes in the spin measurement on the first qubit over an angle $\phi$.)

To test whether a few-parameter model correctly quantifies entanglement if that model is preferred over the FPM by AIC, we estimate a (posterior, Bayesian) probability distribution over the generalized negativities (defined above). We see that the first three curves in
FIG.~\ref{fig_negdist_M1_witness}(a) and the first two curves in FIG.~\ref{fig_negdist_M1_witness}(b), which correspond to the data points above the horizontal line in FIG.~\ref{fig_AIC_M1_witness}, all give consistent estimates of $\mathcal{N}_0$, compared to the actual value of $\mathcal{N}_0$ for the true state (and the same holds for $\mathcal{N}_{1,2}$ (not shown)). Conversely, the estimate cannot be trusted when AIC deems the simple model inferior to the FPM (of course, it may still happen to be a correct estimate, but one could not be sure).
This gives additional evidence for the success  of AIC.
\subsection{Cross modeling for a witness measurement}
We now construct a two-parameter model $\mtwo$ similar in spirit to that discussed for tomographically complete measurements: half the data [on which half the time $(\sigma_x)^{\otimes 4}$ is measured, and half the time $(\sigma_y)^{\otimes 4}$] are used to generate a better model, which is then tested on the other half of the data (also containing both types of measurements equally). We write
\begin{align}
  \rho(\epsilon,q)=&(1-\epsilon)\left[(1-q)\rho_{\rm observation}\right.\nonumber\\
  &\left.+q\ket{\Psi_{\rm target}^{2}(\phi)}\bra{\Psi_{\rm target}^{2}(\phi)}\right]+\epsilon\openone/D.\label{eq_M2_witness}
\end{align}
To find a $\rho_{\rm observation}$---there are many equivalent ones for predicting the outcomes of the witness measurements---we recall that a generic four-qubit state can be expressed as
\begin{align}
  \rho=\sum_{jklm}c_{jklm}\sigma_j\otimes\sigma_k\otimes\sigma_l\otimes\sigma_m,
\end{align}
where $j,k,l,m=1,2,3,4$ where
$\sigma_{1,2,3}$ denote the Pauli matrices $\sigma_{x,y,z}$ and  $\sigma_4=\openone$.  The witness measures the coefficients $c_{jklm}$ where $j,k,l,m$ can be combinations of only 1 and 4 or combinations of only 2 and 4 (\textsl{e.g.}, $c_{1441}$ or $c_{4222}$). We label the $c_{jklm}$'s that can be recovered from witness measurement as $c_{jklm}^w$ ($w$ as in \emph{witness}). We do not include in $c_{jklm}^w$ the coefficient $c_{4444}$, which always equals $1/16$, so that it does not depend on measurement outcomes. We define
 \begin{align}
   \rho_{\rm observation}=\sum_{jklm}c_{jklm}^w\sigma_j\otimes\sigma_k\otimes\sigma_l
\otimes\sigma_m+\openone/16.
 \end{align}

 \begin{figure}[t]
\begin{center}
  \includegraphics[width=165pt]{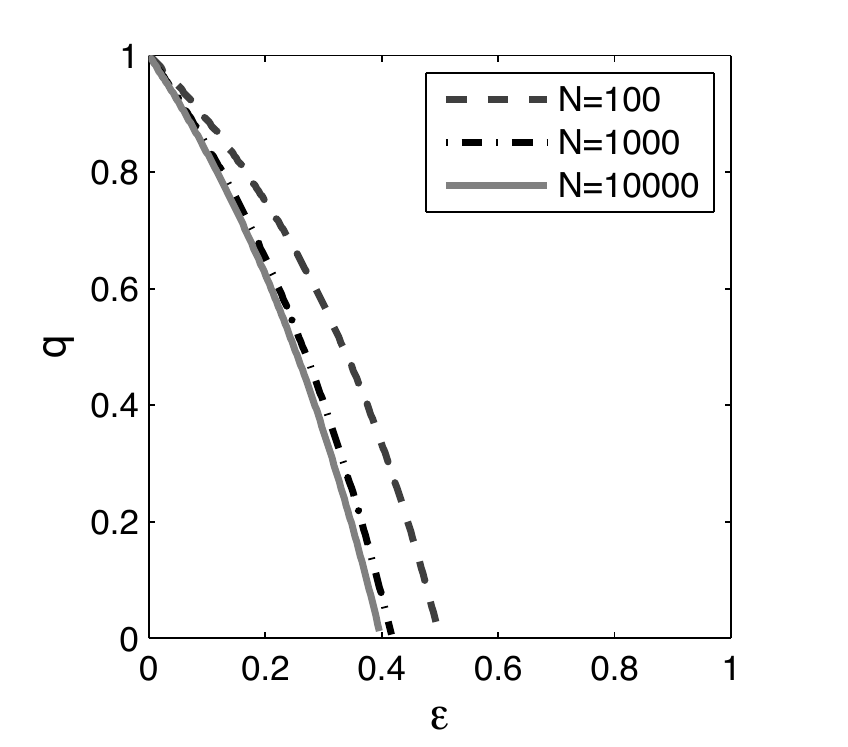}
  \caption{{\em What fraction of the
  model Eq. (\ref{eq_M2_witness}) describes physical states?}: The lower left part separated by the curves is where $\rho(\epsilon,q)$ of Eq. (\ref{eq_M2_witness}) is unphysical  (and so is not actually included in the model), for different number of measurements $N$. }
  \label{fig_unphysical_e_q}
\end{center}
\end{figure}

\begin{figure}[t]
\begin{center}
  \includegraphics[width=195pt]{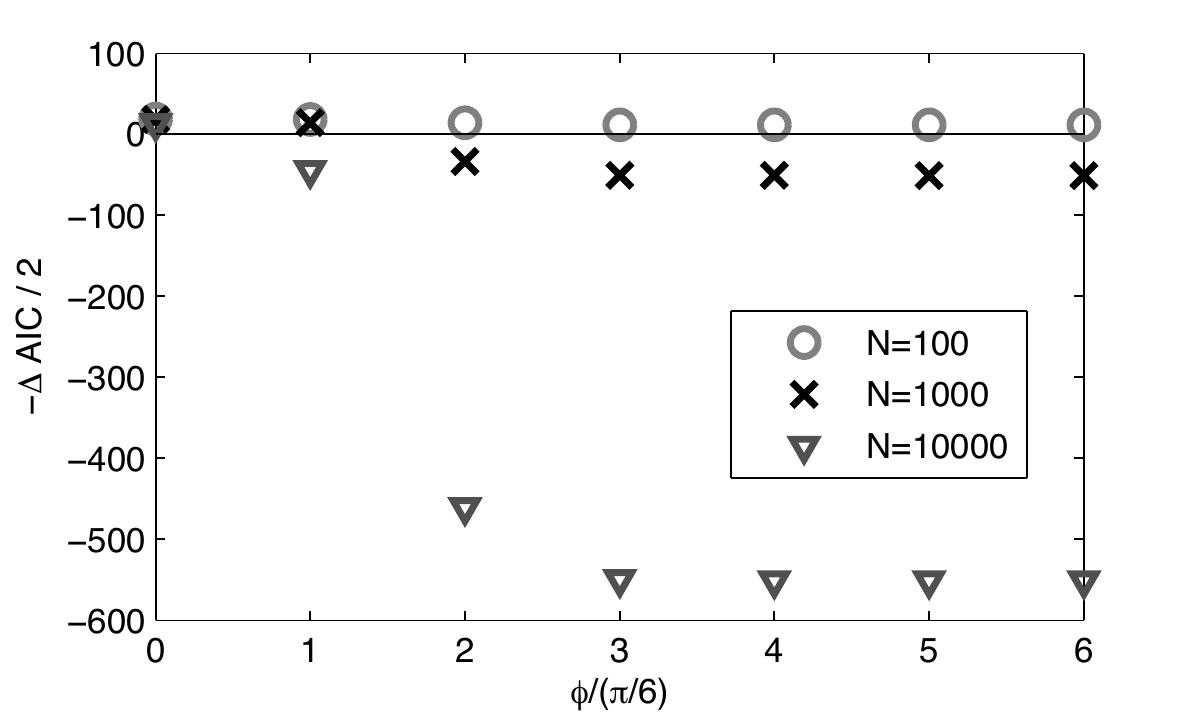}
  \caption{{\em How the AIC ranks our two-parameter model vs. the FPM, for a witness measurement}: The difference between AICs of $\mtwo$ and the FPM, \textsl{i.e.}, $-\Delta{\rm AIC}={\rm AIC(FPM)}-{\rm AIC}(\mtwo)$, for different numbers of witness measurements as a function $\phi$. The target state is $\ket{\Psi_{\rm target}^{2(1)}}$. The horizontal line demarcates $\Delta{\rm AIC}=0$.}
  \label{fig_AIC_M2_witness}
\end{center}
\end{figure}
Note that $\rho_{\rm observation}$ can be considered as a trace-one \emph{pseudostate}, since it is not necessarily positive semi-definite. But the most attractive property of $\rho_{\rm observation}$ is that it preserves the measurement outcomes. It is in fact the unique \emph{pseudostate} that reproduces the exact frequencies of all measurement outcomes {\em and} that has vanishing expectation values for all other {\em un}performed collective Pauli measurements. As a component of $\rho(\epsilon,q)$, we allow $\rho_{\rm observation}$ to be unphysical, but we only keep those $\rho(\epsilon,q)$ that are positive semi-definite.
We checked numerically for what values of $\epsilon$ and $q$ the states end up being physical, and how this  depends on the number of measurements performed.
Physical states are located in the upper right part of the square in FIG.~\ref{fig_unphysical_e_q}. That is, only if $\epsilon$ and/or $q$ are sufficiently large,  so that a sufficiently large amount of $\ket{\Psi_{\rm target}^2}$ and/or $\openone/16$ has been mixed in, does $\rho(\epsilon,q)$ become physical. Depending on the number of measurements, the area of the upper right part is about 69\%-77\% of the whole square. The physical/unphysical boundary shifts closer to the origin as the number of measurements increases.

\begin{figure*}[t]
\begin{center}
  \subfigure[$N=100, \ket{\Psi_{\rm target}^2(\phi)}$]{\includegraphics[width=162pt]{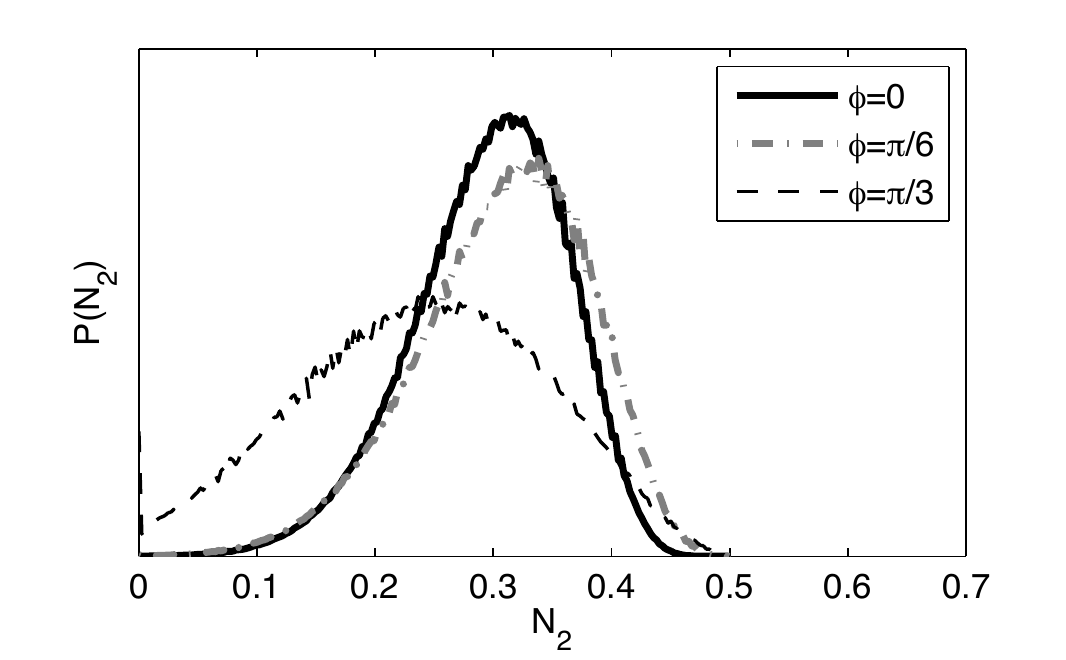}}
  \subfigure[$N=1000, \ket{\Psi_{\rm target}^2(\phi)}$]{\includegraphics[width=162pt]{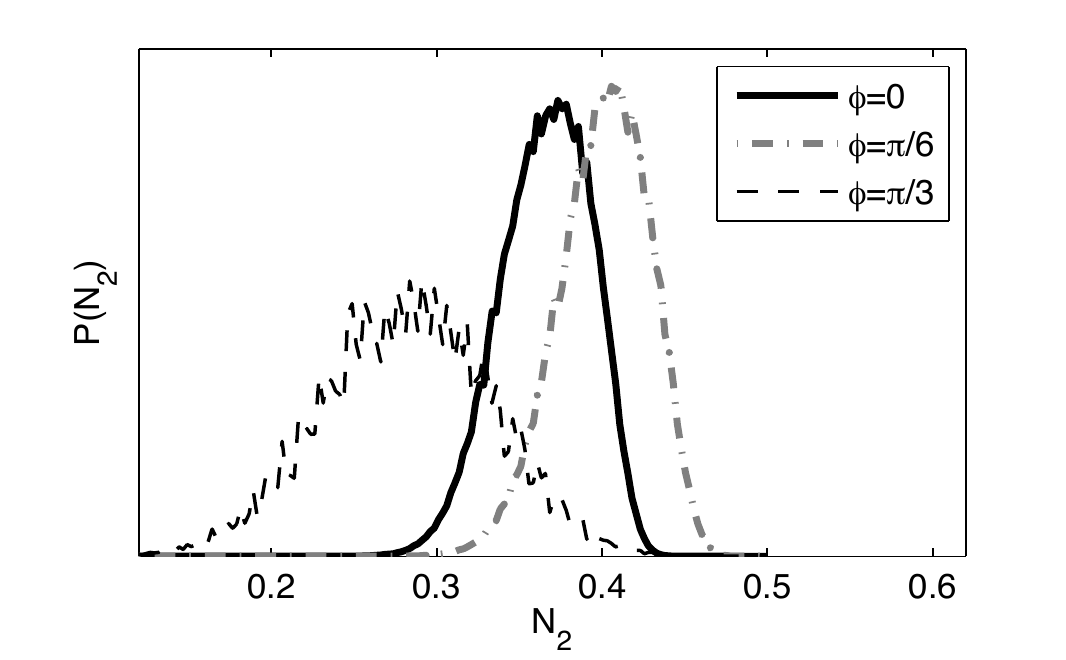}}
  \subfigure[$N=10000, \ket{\Psi_{\rm target}^2(\phi)}$]{\includegraphics[width=162pt]{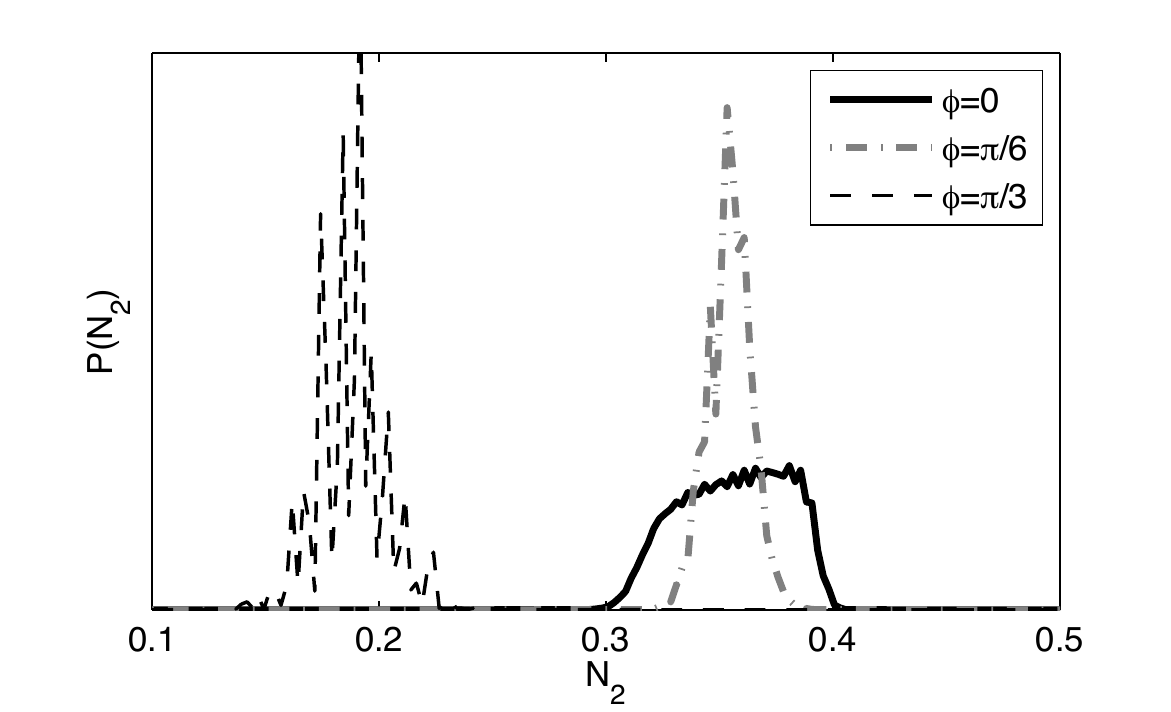}}
  \caption{{\em Quantifying entanglement from a witness measurement}: The posterior distributions for $\mathcal{N}_2$, for different numbers of measurements, using  model $M_{2\phi}$ with $\rho_{\rm observation}$ and target state $\ket{\Psi_{\rm target}^2}$, where $\phi=0, \pi/6, \pi/3$. $\mathcal{N}_2(\rho_{\rm actual})=0.3875$. The same prior is used as in FIG.~\ref{fig_negdist_M1_witness}. Whenever the AIC declares a model superior to the FPM, the estimated entanglement agrees,  within error bars, with the actual value, but may be wrong otherwise. }
  \label{fig_negdist_M2_witness}
\end{center}
\end{figure*}

\begin{figure}[t]
\begin{center}
  \includegraphics[width=195pt]{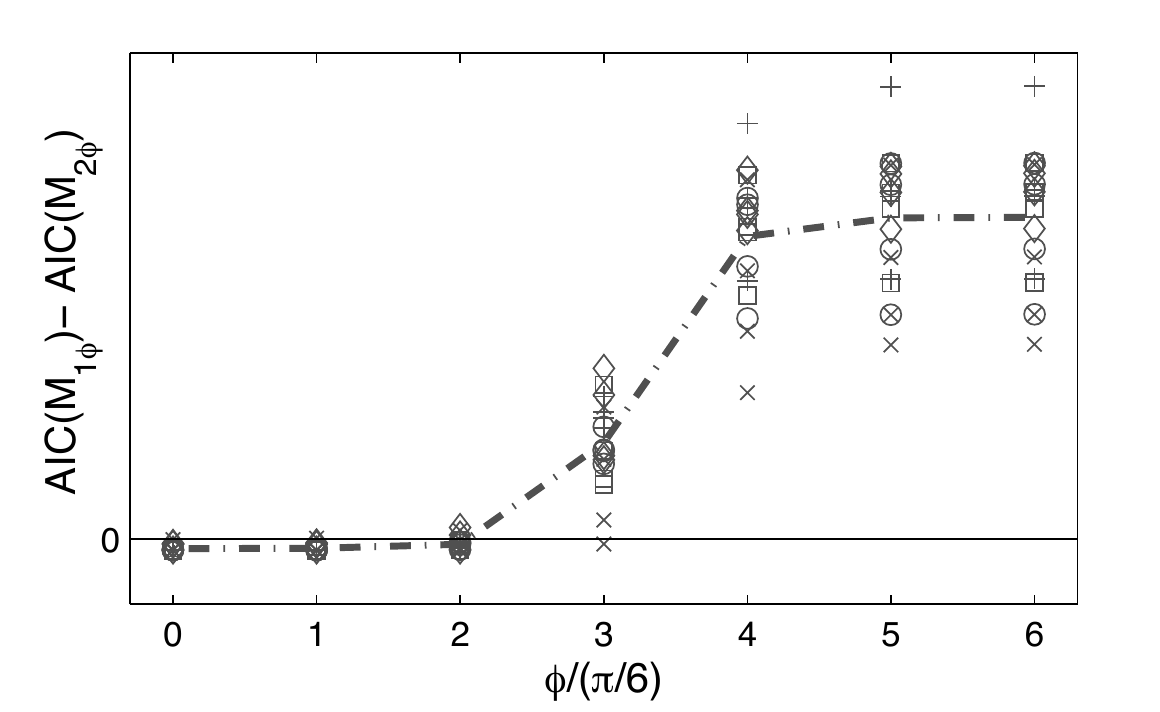}
  \caption{{\em Comparing one- and two-parameter models directly}: The difference between AICs of $\mtwo$ and $\mone$ for 20 different sets of witness measurements ($N=50$) as functions of $\phi$. The target state is $\ket{\Psi_{\rm target}^{2}}$. The horizontal line demarcates $\Delta{\rm AIC}=0$. The dotted-dashed line is the average of all 20 points at each different $\phi$.}
  \label{fig_AIC_M1vsM2_witness}
\end{center}
\end{figure}

We test the two-parameter model (the physical part of it), and show the results in FIG.~\ref{fig_AIC_M2_witness}. We find that for $N=100$ the AIC ranks $\rho_{2\phi}(\epsilon,q)$ better than the FPM, even when the guess about $\phi$ is very imprecise: 100 witness measurements are, unsurprisingly, not enough for a correct reconstruction of the state. When $N=1000$, AIC only prefers the models with a value for $\phi$ within $\pi/6$ of the correct value. And when $N=10000$, the accepted values of $\phi$ are even closer to the true value.

The corresponding posterior distributions of negativities ${\cal N}_2$ are plotted in FIG.~\ref{fig_negdist_M2_witness} for the three better guesses, $\phi=0, \pi/6, \pi/3$.
When $N=100$ all three give decent predictions of $\mathcal{N}_2$ (and indeed, AIC ranks those models highly). For $N=1000$ and $N=10000$, we would only trust the estimates arising from the lower two values of $\phi$, or just the correct value of $\phi$, respectively. This trust is rewarded in FIG.~\ref{fig_AIC_M2_witness}(b) and FIG.~\ref{fig_AIC_M2_witness}(c), as those estimates are indeed correct, within the error bars. In addition, the untrusted estimate for $\phi=\pi/6$ for $N=10000$ still happens to be correct, too.
\subsection{Comparing one- and two-parameter models directly}
Finally, the AIC can compare the one- and two-parameter models $\mone$ and $\mtwo$ directly.
For that purpose one needs to use the {\em same} validation set of data, which implies that the two-parameter model needs {\em additional} data to generate  $\rho_{{\rm observation}}$. Here we display results for just 50 witness measurements, and an additional set of 50 measurements for $\mtwo$. FIG.~\ref{fig_AIC_M1vsM2_witness} shows that even such a small number of additional data is useful if the angle $\phi$ is wrong, and, similarly, it shows that the same small number suffices to detect a wrong single-qubit phase when it is larger than $\pi/3$.

\section{Conclusions}
\label{sec_conclusion}
We applied information criteria, and the Akaike Information Criterion (AIC) developed in Ref.~\cite{A1974} in particular, to quantum state estimation. We showed it to be a powerful method, provided one has a reasonably good idea of what state one's quantum source actually generates.

For each given model, which may include several parameters describing error and noise, as well as some parameters---call them the ideal-state parameters--- describing the state one would like to generate in the ideal (noiseless and error-free) case,  the AIC determines a ranking from the observed data. One can construct multiple models, for instance, models where some ideal-state parameters and some  noise parameters are fixed (possibly determined by previous experiments in the same setup), with others still considered variable.
Crucially, the AIC also easily ranks the full-parameter model (FPM), which uses in principle all exponentially many parameters in the full density matrix, and which is, therefore, the model one would use in full-blown quantum state tomography. This ranking of the FPM can be accomplished without actually having to find the maximum-likelihood state (or its likelihood)---which quickly would run into insurmountable problems for many-qubit systems---by using a straightforward upper bound.

This way, observed data is used to justify {\em a posteriori} the use of the few-parameter models---namely, if the AIC ranks that model above the FPM---and thus our method is in the same spirit as several other recent proposals \cite{GLFBE2010,CPFSGBL-CPL2010} to simplify quantum tomography, by tentatively introducing certain assumptions on the quantum state generated, after which data is used to certify those assumptions (and if the certification fails, one at least knows the initial assumptions were incorrect).

We illustrated the method on (noisy and mis-specified) four-qubit members of the family of Dicke states, and demonstrated its effectiveness and efficiency. For instance, we showed that one can detect mis-specified ideal-state parameters and determine noise and error parameters. We also showed by example the successful application of the method to a specific and useful subtask, that of quantifying multi-qubit entanglement.

\section*{Acknowledgement}
This work was supported by NSF grant PHY-1004219.
\bibliography{hompeschomp}

\end{document}